# Tailed Radio Galaxies as Probes of Cluster Physics in the Square Kilometre Array Era

*Siamak Dehghan\*[1] and Melanie Johnston-Hollitt[1]*

[1]School of Chemical and Physical Sciences, Victoria University of Wellington, PO Box 600, Wellington, 6140, New Zealand, siamak.dehghan@vuw.ac.nz

## Abstract

In recent years, the use of tailed radio galaxies as environmental probes has gained momentum as a method for galaxy cluster detection, examining the dynamics of individual clusters, measuring the density and velocity flows in the intra-cluster medium, and for probing cluster magnetic fields. To date instrumental limitations in terms of resolution and sensitivity have confined this research to the local (z < 0.7) Universe. The advent of SKA-1 surveys however will allow detection of well over 1 million tailed radio galaxies and their associated galaxy clusters out to redshifts of 2 or more. This is in fact ten times more than the current number of known clusters in the Universe. Such a substantial sample of tailed galaxies will provide an invaluable tool not only for detecting clusters, but also for characterizing their intra-cluster medium, magnetic fields and dynamical state as a function of cosmic time. In this paper we present an analysis of the usability of tailed radio galaxies as tracers of dense environments extrapolated from existing deep radio surveys such the Extended Chandra Deep Field-South.

## 1. Introduction

Matter is not uniformly distributed in the Universe, rather it is concentrated in a complex system of tangled galaxy filaments, known as the 'cosmic web'. Where these gigantic filaments intersect, the largest gravitationally bound structures of the Universe, galaxy clusters, are formed. Understanding how these massive systems are created and evolve is crucial to our perception of the large-scale structure of the Universe. In the last few decades numerous methods and sophisticated tools have been introduced or developed in order to detect clusters and other overdense regions of the Universe, and determine their physical characteristics. These methods range from X-ray [1] and Sunyaev-Zel'dovich [2] detections to hierarchical and partitioning clustering analysis of spectroscopic and photometric redshifts [3, 4]. In recent years, several studies on various types of radio sources in the field and cluster environments have demonstrated that some radio sources are remarkable tools to probe the large-scale structures of the Universe.

In this context, the community's attention has recently focused on a class of radio galaxies known as Bent-Tailed sources (BTs). BTs as class encompass all galaxies in which the radio lobes and jets are not aligned linearly with the host galaxy. One subclass of such sources show spectacular structures, in which the jets are bent through large angles. These sources which are known as Narrow-Angle-Tail (NAT) or Head-Tail (HT) radio galaxies, appear to exclusively reside within the dense environment of galaxy clusters. These Fanaroff-Riley class I radio sources [5] are believed to be bent by ram pressure due to their movement in the Intra-Cluster Medium (ICM) [6]. An alternative explanation of the curved radio structure is that buoyancy forces due to density variations of the intra-galactic medium, in which the galaxy is embedded, cause the jets to bend [7]. The coincidence of BTs and clusters in the local ($z \leq 0.07$) Universe, has been found to be at rate of 1-2 per cluster [8]. In addition, there is growing evidence that such association persists up to redshift 2 at the limits of cluster detection [9]. The association between BTs and galaxy clusters, provides an exceptional tool to both trace the large-scale structure of the Universe, and probe the physical conditions within clusters.

## 2. Detecting Tailed Radio Galaxies in the SKA Era

A series of stacked surveys with increasing sensitivity over smaller and smaller areas are planned to be undertaken with the Square Kilometre Array during phase 1 (SKA-1) [10]. This approach, known as a 'wedding cake' survey strategy, will commence with an 'all-sky' survey ($31,000 \, \text{deg}^2$) from 1-2 GHz with $2"$ resolution and 2 µJy sensitivity. Such a survey can be expected to detect numerous BT galaxies and therefore their associated clusters.

In order to assess the capability of SKA-1 surveys to detect BTs, we carried out a detailed radio source examination of the Australia Telescope Large Area Survey of the Chandra Deep Field-South (ATLAS-CDFS) field. The ATLAS-

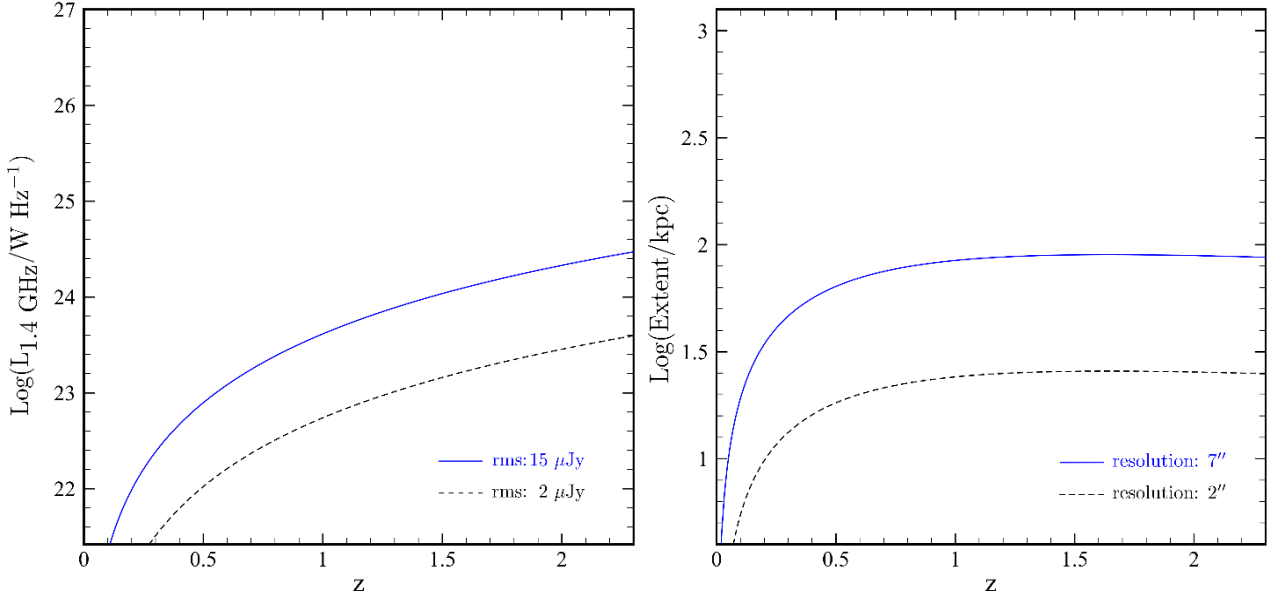

**Figure 1. Left panel:** The blue line and black dashed lines represent the 6σ source detection limit of the ATLAS and SKA-1 surveys with the average sensitivity of 15 and 2 μJy, respectively. **Right panel:** The blue line and black dashed line represent the detection limit of the sources with an extent of 1.5 beams in ATLAS and SKA-1 surveys at 7 & 2 arcseconds resolution, respectively.

CDFS field which spans over $4\,\mathrm{deg}^2$, has been extensively observed with the Australia Telescope Compact Array (ATCA) at 1.4 GHz [11]. The radio image has a resolution of about $17'' \times 7''$ and reaches a depth of 15 μJy, allowing the exploration of a significantly fainter population of sources, which are expected to be readily detected by the SKA-1 surveys.

The ATLAS data were supplemented by a $2''$ resolution 1.4 GHz Very Large Array (VLA) image of the Extended CDFS (ECDFS) with depth of 10 μJy/beam [12]. Note that the capabilities of the proposed SKA-1 all-sky survey are comparable in resolution with the VLA-ECDFS survey, though the SKA-1 survey will be slightly deeper (see Figure 1).

The images of the CDFS field were visually inspected for low-surface-brightness radio sources. As a result, 56 such radio sources were identified and classified out of a total of ~ 3000. These sources include 45 BT galaxies and a radio relic candidate. The remaining diffuse radio sources could not be unambiguously classified. Based on extrapolating from these results, future all sky 1.4 GHz continuum surveys comparable to the ATLAS survey in terms of resolution and sensitivity such as EMU [13], will detect at least 590,000 extended low-surface-brightness radio sources, including over 470,000 BT radio galaxies. Advancing to SKA-1 surveys should see this double to over 1 million BT detections. Such extensive samples of tailed radio sources will significantly expand our knowledge of the origins and evolution of BTs and the mutual correlation between these sources and the environment in which they reside.

By using BT galaxies as tracers of clusters, the SKA-1 may detect over 1 million clusters. This is more than twenty times the number of clusters currently known, and will be comparable with or exceed over the number expected to be detected with next generation X-ray instruments such as eROSITA [14]. If we presume the same clusters are detected by both eROSITA and SKA-1 surveys, we will be in an unprecedented position to explore the multi-wavelength properties of clusters from the present to a redshift of 2.

### 3. Tailed Radio Galaxies as Environmental Probes

The morphology of BTs is primarily a function of the environment that they have resided in [15]. In particular, there are some significant environmental effects on the morphology of tailed radio galaxies, both on local and large scales. These are:

- A dense ICM that generates strong ram pressure and buoyancy forces on the plumes of tailed radio galaxies, especially NATs. The ram pressure is either the result of the host galaxy falling into the gravitational centre of the cluster, or the ICM blowing past the galaxy like a wind in a cluster weather system.

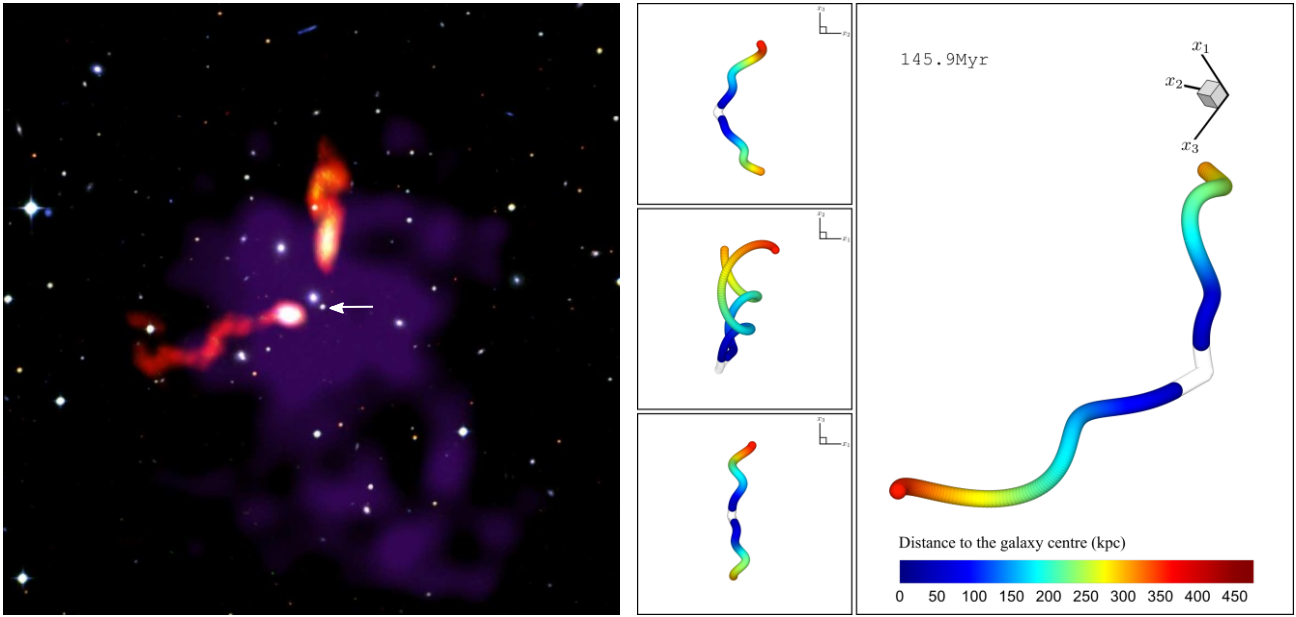

**Figure 2. Left panel:** Three colour (blue, red, and infrared) image of A3135 overlaid with Chandra observations (purple) and 1.4 GHz ATCA observations of the PKS J0334-3900 radio galaxy. The image colour scales have been adjusted in a similar manner to images made for the Hubble Heritage Project. The close companion galaxy is shown by a white arrow. **Right panel:** Multi-view orthographic projections of the PKS-J0334-3900 model. The transparent region corresponds to the gap in the radio image close to the AGN. The top/bottom jets are moving away/towards us respectively (Figures adapted from [16]).

- The radio structures of tailed radio galaxies are dramatically altered by the dynamics of large-scale structures. Mergers, mass accretions, and collapse of clusters will bend the jets of BTs and NATs.
- The local environment of host galaxies may cause more complex radio structure in tailed radio galaxies. Gravitational interactions, including bound and unbound orbital motions, and near-miss passes affect the structure of the tails.

In order to understand the role of the variety of factors that influence the radio morphology, we constructed a simple model that generates the overall radio structure of the sources in different habitats. This simple model was applied to simulate a prominent BT galaxy in the Abell 3135 galaxy cluster (see Figure 2). As a result in this case, we found that a combination of orbital and precessional motions along with a cluster wind were required to generate the observed morphology of this tailed radio galaxy [16]. However, the model is capable of fully disentangling different environmental effects on BTs.

Using data and modelling we are also able to constrain the magnetic field of the cluster, demonstrating BTs as valuable sources in the exploration of cosmic magnetism. Thus, the anatomy of tailed radio galaxies is an invaluable source of environmental information, in which a history of the past interactions, such as complex galaxy motions and cluster merger shocks are remarkably preserved. Such procedures carried out on a sample of BTs will provide strong constraints on the evolution and dynamical state of their host clusters as a function of redshift.

## 4. Conclusions

While a large variety of methods and techniques has been developed and deployed to detect the overdense regions of the Universe and examine their physical characteristics, future radio observatories such as SKA-1 will provide a more comprehensive, deeper, and yet observationally cheaper approach via all-sky continuum surveys. The unprecedented capabilities of such surveys will allow us to readily detect more than 1 million diffuse extended radio sources. Systematic studies of these extended radio sources in the near future, will boost understanding of the role of environment on the structures of extragalactic radio sources and evolution of extended radio sources, as well as characteristics of galaxy clusters, including their magnetic fields strength, dynamical state, and distribution. The resulting discoveries will not only benefit the radio astronomy community, but will also provide a proving ground for a range of physical and cosmological theories.